\documentclass[reprint, 10pt, amsmath, amssymb, floatfix, aps, prd, superscriptaddress, nofootinbib, nobibnotes,onecolumn]{revtex4-1}

\pdfoutput=1
\usepackage{graphicx}
\graphicspath{{figures/}}
\usepackage[utf8]{inputenc}
\usepackage{amsmath}
\usepackage{amsfonts}
\usepackage{amssymb}
\usepackage{multirow}
\usepackage{CJKutf8}
\usepackage{color}
\usepackage[colorlinks=true, linkcolor=red, citecolor=blue]{hyperref}
\usepackage[capitalise]{cleveref}
\usepackage{acro}
\usepackage{adjustbox}
\usepackage{array}
\usepackage{natbib}
\usepackage{dblfnote}
\DFNalwaysdouble
\usepackage{slashed}
\usepackage{dcolumn}
\usepackage{hyperref}
\usepackage{geometry}
\usepackage{inputenc}

\def\({\left(}
\def\){\right)}
\def\[{\left[}
\def\]{\right]}
\def\be{\begin{eqnarray}}
\def\ee{\end{eqnarray}}

\DeclareAcronym{GW}{
  short = GW ,
  long = gravitational wave ,
  short-plural = s 
}
\DeclareAcronym{LIGO}{
  short = LIGO ,
  long = Laser Interferometer Gravitational-wave Observatory ,
  short-plural = 
}
\DeclareAcronym{LISA}{
  short = LISA ,
  long = Laser Interferometer Space Antenna ,
  short-plural =  
}
\DeclareAcronym{SKA}{
  short = SKA ,
  long = Square Kilometre Array ,
  short-plural =  
}
\DeclareAcronym{FLRW}{
  short = FLRW ,
  long = Friedmann-Lemaitre-Robertson-Walker ,
  short-plural =  
}

\begin{document}

\title{On the Gauge Invariance of Scalar Induced Gravitational Waves: Gauge Fixings Considered}

\author{Zhe Chang}
\email{changz@ihep.ac.cn}
\affiliation{Theoretical Physics Division, Institute of High Energy Physics, Chinese Academy of Sciences, Beijing 100049, People's Republic of China}
\affiliation{School of Physical Sciences, University of Chinese Academy of Sciences, Beijing 100049, People's Republic of China}
\author{Sai Wang}
\email{wangsai@ihep.ac.cn}
\affiliation{Theoretical Physics Division, Institute of High Energy Physics, Chinese Academy of Sciences, Beijing 100049, People's Republic of China}
\affiliation{School of Physical Sciences, University of Chinese Academy of Sciences, Beijing 100049, People's Republic of China}
\author{Qing-Hua Zhu}
\email{zhuqh@ihep.ac.cn}
\affiliation{Theoretical Physics Division, Institute of High Energy Physics, Chinese Academy of Sciences, Beijing 100049, People's Republic of China}
\affiliation{School of Physical Sciences, University of Chinese Academy of Sciences, Beijing 100049, People's Republic of China}


\begin{abstract}
The energy density spectrum is an observable of gravitational waves. Divergence has appeared in the energy density spectra of the scalar induced gravitational waves for different gauge fixings. To resolve the discrepancy, we investigate the gauge invariance of the scalar induced gravitational waves. It is shown that the gauge invariant induced gravitational waves can be obtained by subtracting the fictitious tensor perturbations via introducing the counter term composed of the first order scalar perturbations. The kernel function uniquely determines the energy density spectrum of the scalar induced gravitational waves. We explicitly calculate the gauge invariant kernel functions in the Newtonian gauge and the uniform density gauge, respectively. The discrepancy between the energy density spectra upon the Newtonian gauge and the uniform density gauge is shown to be eliminated in the gauge invariant framework. In fact, the gauge invariant approach is also available to other kinds of gauge fixings.
\end{abstract}

\maketitle
\acresetall

\section{Introduction}\label{sec:intro}
Gravitational waves are expected to be produced from the scalar perturbations via the non-linear couplings in the second or higher order cosmological perturbation theory \cite{Mollerach:2003nq,Ananda:2006af,Baumann:2007zm,Assadullahi:2009jc}. 
Such a kind of gravitational waves are called the induced gravitational waves. 
To measure the gravitational waves, one defines the energy density spectrum to be the physical observable \cite{Allen:1997ad}. 
For the induced gravitational waves, however, the energy density spectrum has been shown to be divergent across the literature \cite{Inomata:2019yww,Yuan:2019fwv,DeLuca:2019ufz,Lu:2020diy,Tomikawa:2019tvi,Hwang:2017oxa,Giovannini:2020qta,Ali:2020sfw}. 
To subtract the fictitious perturbations, it has been suggested that the gravitational waves should be studied in the gauge invariant frame \cite{Domenech:2017ems,Gong:2019mui,Bruni:1996im,Matarrese:1997ay,Malik:2008im,Nakamura:2006rk,Chang:2020tji}.
This frame was further suggested to be synchronous \cite{DeLuca:2019ufz}, since the measurements are performed in the synchronous gauge \cite{Maggiore:1999vm}. 
Most recently, we proposed such a well-defined gauge invariant synchronous frame \cite{Chang:2020iji}, by following the Lie derivative method \cite{Matarrese:1997ay,Nakamura:2006rk,Malik:2008im,Bruni:1996im,Chang:2020tji}. 
Meanwhile, we calculated the energy density spectrum of the induced gravitational waves in such a framework. 

In this work, by following the method proposed in Ref.~\cite{Chang:2020iji}, we will explicitly show how to resolve the discrepancy between the energy density spectra calculated in the Newtonian gauge and the uniform density gauge. 
In fact, this discrepancy has been studied in previous works \cite{Lu:2020diy,Ali:2020sfw}. 
The energy density spectrum at late time was found to increase as $\eta^{6}$ in the uniform density gauge, compared with the Newtonian gauge. 
Here, $\eta$ denotes a conformal time of the Universe. 
When $\eta\rightarrow\infty$, this strange result might imply a break down of the cosmological perturbation theory. 
However, in the gauge invariant framework, the infinity is expected to be eliminated by considering a counter term, which is composed of the first order scalar perturbations \cite{Chang:2020tji,Chang:2020iji}. 
Note that we consider only the first order scalar perturbations in this work. 
In fact, our method is also available when we take into account the first order vector and tensor perturbations.

The remainder of the paper is arranged as follows. 
In Section \ref{sec:gi}, we introduce the gauge invariant induced gravitational waves as well as the kernel function. 
In Section \ref{sec:new} and \ref{sec:ud}, we explicitly calculate the kernel functions in the Newtonian gauge and the uniform density gauge, respectively. 
In Section \ref{sec:kernel}, we compare the kernel functions. 
The conclusion and discussion are summarized in Section \ref{sec:conc}.

\section{Gauge invariant induced gravitational waves}\label{sec:gi}

In the \ac{FLRW} spacetime, the background metric is given by
\begin{eqnarray}
  g_{\mu \nu}^{(0)} {\rm{d}} x^{\mu} {\rm{d}} x^{\nu} & = & a^2 (\eta) (- {\rm{d}}
  \eta^2 + \delta_{i   j} {\rm{d}} x^i {\rm{d}} x^j)\ , 
\end{eqnarray}
where $a (\eta)$ is a scale factor of the Universe at $\eta$. 
To study the scalar induced gravitational waves, we consider the metric perturbations up to second order, i.e., \cite{Mollerach:2003nq,Ananda:2006af,Baumann:2007zm,Assadullahi:2009jc}
\begin{eqnarray}
  {\rm{d}} s^2 & = & a^2 \left( - \(1 + 2 \phi^{(1)}\) {\rm{d}} \eta^2 + 2
  \partial_i b^{(1)} {\rm{d}} \eta {\rm{d}} x^i + \left( \(1 - 2 \psi^{(1)}\)
  \delta_{i   j} + 2 \partial_i \partial_j e^{(1)} + \frac{1}{2} h_{i
    j}^{(2)} \right) {\rm{d}} x^i {\rm{d}} x^j \right)\ ,  \label{2}
\end{eqnarray}
where we have neglected the first order vector and tensor perturbations \cite{Chang:2020tji}. 
For simplicity, we make conventions for the scalar perturbations $\phi \equiv \phi^{(1)}$, $b \equiv b^{(1)}$, $\psi \equiv \psi^{(1)}$, $e \equiv  e^{(1)}$, and the tensor perturbations $h_{i   j} \equiv h^{(2)}_{i   j}$. 
The transverse and traceless conditions are satisfied, i.e., $\delta^{j   l} \partial_l h_{i   j} = 0$ and $\delta^{i j} h_{i  j} = 0$.

The gauge invariant second order tensor perturbations $H_{i   j}$ are defined as \cite{Chang:2020iji}
\be
  H_{i   j} = h_{i   j} - \Lambda^{  l   m}_{i
    j} \mathcal{X}_{l   m}\ , \label{3}
\ee
where second term at the right hand side of the above equation is shown to be
\be
\Lambda^{  l   m}_{i
	j} \mathcal{X}_{l   m} = \Lambda^{  l   m}_{i
	j} \big(8 e \partial_l \partial_m \phi + 2\partial^s e \partial_s \partial_l \partial_m e -2(\partial_0 e- b)\partial_l \partial_m (\partial_0 e- b)  \big) \label{5}
\ee
Here, the $\Lambda^{l   m}_{i   j}$ denotes a transverse and traceless operator. Explicit expression of $\mathcal{X}_{i j}$ was given in Refs.~\cite{Chang:2020iji,Chang:2020tji}.
It should be noticed that $\mathcal{X}_{kl}$ is uniquely determined by (the square of) the first order scalar perturbations. 
Since $h_{i   j}$ are gauge dependent, the second term at the right hand side of Eq.~(\ref{3}) can be understood as the gauge dependent counter term that ensures the gauge invariance of $H_{i j}$. 
Therefore, $H_{i   j}$ are obtained via subtracting the fictitious components in $h_{i   j}$.

We briefly review the well-known results on $h_{i j}$ as the first step. 
The equation of motion of $h_{i j}$ is given by \cite{Mollerach:2003nq,Ananda:2006af,Baumann:2007zm,Assadullahi:2009jc} 
\begin{eqnarray}
  \partial_0^2 h_{i   j} + 2\mathcal{H} \partial_0 h_{i   j} -
  \Delta h_{i   j} & = & - 4 \Lambda^{l   m}_{i   j}
  \mathcal{S}_{l   m}\ , \label{6}
\end{eqnarray}
where $\mathcal{H}$ denotes the conformal Hubble parameter, and $\mathcal{S}_{l   m  }$ is a source term depending on the gauge fixings. 
The Fourier mode of $\Lambda^{l   m}_{i   j}$ can be expressed in terms of the polarization tensors $\varepsilon^{\lambda}_{i   j}$ ($\lambda = +, \times$), namely, \cite{Chang:2020tji}
\begin{equation}
  \Lambda^{  l   m}_{i   j} \left( \textbf{k} \right) =
  \delta_{\lambda \bar{\lambda}} \varepsilon^{\lambda}_{i   j} \left(
  \textbf{k} \right) \varepsilon^{\bar{\lambda}, l   m} \left( \textbf{k}
  \right)\ , 
\end{equation}
where the polarization tensors satisfy the condition $\varepsilon_{i j}^{\lambda} \varepsilon^{\bar{\lambda}, i   j} = \delta^{\lambda\bar{\lambda}}$. 
Eq.~(\ref{6}) can be rewritten in the momentum space, i.e., 
\begin{eqnarray}
  \left( h_{\textbf{k}}^{\lambda} \right)'' + 2\mathcal{H}   \left(
  h_{\textbf{k}}^{\lambda} \right)' + k^2 h_{\textbf{k}}^{\lambda} & = & 4
    \mathcal{S}_{\textbf{k}}^{\lambda}\ ,\label{8}
\end{eqnarray}
where the prime stands for a derivative with respect to $\eta$, and we denote $h_{\textbf{k}}^{\lambda} \equiv \varepsilon^{\lambda, i j} h_{\textbf{k}, i   j}$ and $\mathcal{S}_{\textbf{k}}^{\lambda} \equiv - \varepsilon^{\lambda, i j} \mathcal{S}_{\textbf{k}, i   j}$ for simplicity. 
Here, $h_{\textbf{k}, i  j}$ and $\mathcal{S}_{\textbf{k}, i   j}$ are the Fourier modes of $h_{ij}$ and $\mathcal{S}_{i   j}$, respectively. 
 $\mathcal{S}_{\textbf{k}}^{\lambda}$ is given by 
\begin{eqnarray}
  \mathcal{S}_{\textbf{k}}^{\lambda} & = & \int \frac{{\rm{d}}^3 p}{\sqrt{(2
  \pi)^3}} \left\{ \varepsilon^{\lambda, ij} (k) p_i p_j f \left(
  \left| \textbf{k} - \textbf{p} \right|, p, \eta \right) \Phi_{\textbf{k} -
  \textbf{p}} \Phi_{\textbf{p}} \right\}\ ,  \label{9}
\end{eqnarray}
where $f \left( \left| \textbf{k} - \textbf{p} \right|, p, \eta \right)$ is a transfer function dependent on the gauge fixings, $\Phi_{\textbf{p}} = {3 (1 + w)}/{(5 + 3 w)} \zeta_{\textbf{p}}$, and $\zeta_{\textbf{p}}$ the primordial
curvature perturbations. 
Following Green's function method, we obtain 
\begin{eqnarray}
  h_{\textbf{k}}^{\lambda} & = & 4 \int_0^{\eta} {\rm{d}} \bar{\eta} \left\{
  \frac{a (\bar{\eta})}{a (\eta)} G_{\textbf{k}} (\eta, \bar{\eta})
  S_{\textbf{k}} (\bar{\eta}) \right\} \nonumber\\
  & \equiv & \int \frac{{\rm{d}}^3 p}{\sqrt{(2 \pi)^3}} \left\{
  \varepsilon^{\lambda, i   j} (k) p_i p_j \Phi_{\textbf{k} -
  \textbf{p}} \Phi_{\textbf{p}} I_h \left( \left| \textbf{k} - \textbf{p}
  \right|, p, \eta \right) \right\}\ , \label{10}
\end{eqnarray}
where $G_{\textbf{k}} (\eta, \bar{\eta})$ denotes the Green's function, and the kernel function $I_{h}$ is defined as
\begin{eqnarray}
  I_h \left( \left| \textbf{k} - \textbf{p} \right|, p, \eta \right) & = &
  \frac{4}{k^2} \int^{\eta}_0 {\rm{d}} (k \bar{\eta}) \left\{ \frac{a
  (\bar{\eta})}{a (\eta)} k   G_{\textbf{k}} (\eta, \bar{\eta}) f
  \left( \left| \textbf{k} - \textbf{p} \right|, p, \bar{\eta} \right)
  \right\}\ . \label{11}
\end{eqnarray}
Eqs.~(\ref{6})--(\ref{11}) have been derived in the semi-analytic way \cite{Kohri:2018awv}.
Moreover, we will derive the explicit expressions of $f\left( \left| \textbf{k} - \textbf{p} \right|, p, \bar{\eta} \right)$, given two different gauge fixings in the following sections. 

We denote $\mathcal{X}_{\textbf{k}, l m}$ as the Fourier mode of the gauge dependent counter term $\mathcal{X}_{l m}$.
Based on Eq.~(\ref{5}), $\mathcal{X}_{\textbf{k}, l m}$ is obtained to be 
\begin{eqnarray}
  \varepsilon^{\lambda, l   m} \left( \textbf{k} \right)
  \mathcal{X}_{\textbf{k}, l   m} & = & \int \frac{{\rm{d}}^3 p}{\sqrt{(2
  \pi)^3}} \left\{ \varepsilon^{\lambda, l   m} \left( \textbf{k}
  \right) p_l p_m \Phi_{\textbf{k} - \textbf{p}} \Phi_{\textbf{p}} I_{\chi}
  \left( \left| \textbf{k} - \textbf{p} \right|, p, \eta \right) \right\}\ ,  \label{14}
\end{eqnarray}
where the kernel function $I_{\chi} \left( \left| \textbf{k} - \textbf{p} \right|, p, \eta \right)$ takes the form of
\begin{eqnarray}
	I_{\chi} \left( \left| \textbf{k} - \textbf{p} \right|, p, \eta \right) & = & - \frac{4 }{\left| \textbf{k} - \textbf{p} \right|^2}
	T_e \left( \left| \textbf{k} - \textbf{p} \right| \eta \right) T_{\psi} (p
	\eta) - \frac{4}{p^2} T_{\psi} \left( \left| \textbf{k} - \textbf{p} \right|
	\eta \right) T_e (p \eta) \nonumber\\
	&  & + 2 \left( \frac{1}{\left| \textbf{k} - \textbf{p} \right|^2} T_e'
	\left( \left| \textbf{k} - \textbf{p} \right| \eta \right) - \frac{1}{\left|
		\textbf{k} - \textbf{p} \right|} T_b \left( \left| \textbf{k} - \textbf{p}
	\right| \eta \right) \right) \left( \frac{1}{p^2} T_e' (p \eta) -
	\frac{1}{p} T_b (p \eta) \right)\nonumber\\
	&  & + 2 \left( \frac{(\textbf{k} - \textbf{p}) \cdot \textbf{p}}{\left|
		\textbf{k} - \textbf{p} \right|^2 p^2} \right)  T_e \left( \left|
	\textbf{k} - \textbf{p} \right| \eta \right) T_e (p \eta), \label{15}
\end{eqnarray}
and $T_{s} (k \eta)$($s = e, \psi, b, \phi$) denote the transfer functions of
the first order scalar perturbations, i.e.,
\begin{eqnarray}
  k^2 e_k (\eta) & = & \Phi_k T_e (k   \eta)\ , \\
  \psi_k (\eta) & = & \Phi_k T_{\psi} (k \eta)\ , \\
  k   b_k (\eta) & = & \Phi_k T_b (k \eta)\ , \\
  \phi_k (\eta) & = & \Phi_k T_{\phi} (k \eta)\ . 
\end{eqnarray}
Here, $e_k$, $\psi_k$, $b_k$ and $\phi_k$ denote the Fourier modes of $e$, $\psi$, $b$ and $\phi$, respectively. 
They are obtained via solving the first order Einstein field equations.

To study the gauge invariant second order gravitational waves $H_{i j}$, we also express Eq.~(\ref{3}) in the momentum space, namely,
\begin{eqnarray}
  H_{\textbf{k}}^{\lambda} & = & h_{\textbf{k}}^{\lambda} -
  \varepsilon^{\lambda, l   m} \mathcal{X}_{\textbf{k}, l   m}\ , \label{12}
\end{eqnarray}
where $H_{\textbf{k}}^{\lambda} \equiv \varepsilon^{\lambda, i   j}
H_{\textbf{k}, i   j}$, and $H_{\textbf{k}, i   j}$ is the Fourier mode of $H_{i j}$. 
Similarly, we define $H_{\textbf{k}}^{\lambda}$ in terms of a kernel function $I_H$, namely, 
\begin{eqnarray}
  H_{\textbf{k}}^{\lambda} & = & \int \frac{{\rm{d}}^3 p}{\sqrt{(2 \pi)^3}}
  \left\{ \varepsilon^{\lambda}_{i   j} (k) p_i p_j \Phi_{\textbf{k} -
  \textbf{p}} \Phi_{\textbf{p}} I_H \left( \left| \textbf{k} - \textbf{p}
  \right|, p, \eta \right) \right\}\ . \label{13}
\end{eqnarray}
Considering Eqs.~(\ref{11}) and (\ref{14}), we can obtain a simple relation among the three kernel functions, i.e.,
\begin{equation}
  I_H \left( \left| \textbf{k} - \textbf{p} \right|, p, \eta \right) = I_h
  \left( \left| \textbf{k} - \textbf{p} \right|, p, \eta \right) - I_{\chi}
  \left( \left| \textbf{k} - \textbf{p} \right|, p, \eta \right)\ . \label{20}
\end{equation}
Both of $I_h$ and $I_{\chi}$ are gauge dependent. However, we show that the difference between $I_h$ and $I_{\chi}$ is the gauge invariant kernel function $I_H$. 
The energy density spectrum ($\Omega_{\rm{GW}}$) of the scalar induced gravitational waves is uniquely determined by the kernel function \cite{Kohri:2018awv,Wang:2019kaf}.
Therefore, it is not doubted that $\Omega_{\rm{GW}}$ defined with $I_H$ is also gauge invariant. 
In the following sections, we will consider two gauge fixings, i.e., the Newtonian gauge and the uniform density gauge, for the metric perturbations in Eq.~(\ref{2}). 
We will show that the kernel function $I_H$ remains the same in both of the two gauge fixings. 
For simplicity, we consider only the radiation dominated epoch, implying $w=c_{s}^{2}=1/3$ \cite{Mukhanov:1990me}.

\section{Kernel functions in the Newtonian gauge}\label{sec:new}

In this section, we consider the metric perturbations up to second order in the Newtonian gauge, i.e.,
\begin{eqnarray}
  {\rm{d}} s^2 & = & a^2 \left( - (1 + 2 \phi) {\rm{d}} \eta^2 + \left( (1 - 2
  \psi) \delta_{i   j} + \frac{1}{2} h_{i   j} \right) {\rm{d}}
  x^i {\rm{d}} x^j \right)\ . 
\end{eqnarray}
Since the second order scalar perturbations do not contribute to the equation of motion of the scalar induced gravitational waves \cite{Chang:2020tji}, we just consider the scalar perturbations up to first order. 
Similarly, the energy momentum tensor up to first order is given by
\begin{eqnarray}
  T_{00} & = & a^2 (1 + 2 \phi) \rho^{(0)} + a^2 \rho^{(1)}\ , \\
  T_{0 i} & = & - a^2 (P^{(0)} + \rho^{(0)}) \upsilon^{(1)}_i\ , \\
  T_{i   j} & = & a^2 \delta_{i   j} ((1 - 2 \psi) P^{(0)} +
  P^{(1)})\ , 
\end{eqnarray}
where $\rho^{(0)}$ and $P^{(0)}$ denote the background density and pressure, respectively, and $\rho^{(1)}$,  $P^{(1)}$ and $\upsilon_i^{(1)}$ denote the first order density, pressure, and velocity perturbations, respectively.

Based on the first order Einstein field equations, the equations of motion of the first order scalar perturbations are given by \cite{Mukhanov:1990me}
\begin{eqnarray}
  3 \partial_0^2 \psi - \Delta \psi + 3\mathcal{H} (\partial_0 \phi + 3
  \partial_0 \psi) & = & 0\ , \\
  \psi - \phi & = & 0\ . 
\end{eqnarray}
In the momentum space, we rearrange the above two equations in the form of 
\begin{eqnarray}
  3 x \partial_x^2 \psi_k (x) + 12 \partial_x \psi_k (x) + x \psi_k (x) & = &
  0\ , \label{27}\\
  \phi_k (x) & = & \psi_k (x)\ , \label{28}
\end{eqnarray}
where we let $x \equiv k \eta$. 
Here, we utilize $\mathcal{H}\propto 1/\eta$ in the derivation. 
By solving Eqs.~(\ref{27}) and (\ref{28}), we obtain the transfer functions of $\psi_k$ and $\phi_k$ as 
\begin{eqnarray}
  T_{\phi} (x) = T_{\psi} (x) & = & \frac{9}{x^2} \left( \frac{\sqrt{3}}{x}
  \sin \left( \frac{x}{\sqrt{3}} \right) - \cos \left( \frac{x}{\sqrt{3}}
  \right) \right)\ . 
\end{eqnarray}

Based on the second order Einstein field equations, we obtain the source term at the right hand side of Eq.~(\ref{6}), namely,
\begin{eqnarray}
  \Lambda^{i   j}_{l   m} \mathcal{S}_{i   j} & = &
  \Lambda^{i   j}_{l   m} \left( 3 \phi \partial_i \partial_j
  \phi + \frac{2}{\mathcal{H}} \partial_0 \phi \partial_i \partial_j \phi +
  \frac{1}{\mathcal{H}^2} \partial_0 \phi \partial_i \partial_j \partial_0
  \phi \right)\ . \label{30}
\end{eqnarray}
Following Eqs.~(\ref{9}) and (\ref{30}), we obtain the transfer function $f \left( \left| \textbf{k} - \textbf{p} \right|, p, \eta \right)$ to be 
\begin{eqnarray}
  f \left( \left| \textbf{k} - \textbf{p} \right|, p, \eta \right) & = & 2
  T_{\psi} \left( \left| \textbf{k} - \textbf{p} \right| \eta \right) T_{\psi}
  (p \eta) + \left( \eta T'_{\psi} \left( \left| \textbf{k} - \textbf{p}
  \right| \eta \right) + T_{\psi} \left( \left| \textbf{k} - \textbf{p}
  \right| \eta \right) \right) (\eta T'_{\psi} (p \eta) + T_{\psi} (p \eta))\ . \label{31}
\end{eqnarray}
On the other side, the counter term $\mathcal{X}_{l   m}$ is shown to vanish in the Newtonian gauge, i.e., $\mathcal{X}_{l   m} = 0$ \cite{Chang:2020iji}.
This implies
\begin{eqnarray}
  I_{\chi} \left( \left| \textbf{k} - \textbf{p} \right|, p, \eta \right) & = & 0\ . 
\end{eqnarray}
Therefore, the gauge invariant kernel function of $I_H$ takes the same value as $I_h$ in Eq.~(\ref{11}), namely, 
\begin{eqnarray}
  I_H \left( \left| \textbf{k} - \textbf{p} \right|, p, \eta \right) & = &
  \frac{4}{k^2} \int^{\eta}_0 {\rm{d}} (k \bar{\eta}) \left\{ \frac{a
  (\bar{\eta})}{a (\eta)} k   G_{\textbf{k}} (\eta, \bar{\eta}) f
  \left( \left| \textbf{k} - \textbf{p} \right|, p, \bar{\eta} \right)
  \right\}\ , 
\end{eqnarray}
where $f \left( \left| \textbf{k} - \textbf{p} \right|, p, \bar{\eta} \right)$ is given in Eq.~(\ref{31}), and $k   G_k (\eta, \bar{\eta}) = \sin (k (\eta - \bar{\eta}))$ in the radiation dominated epoch.
Since $\Omega_{\mathrm{GW}}$ is uniquely determined by the kernel function, the energy density spectrum calculated in the gauge invariant framework takes the same as the one calculated in the Newtonian gauge.

\section{Kernel functions in the uniform density gauge}\label{sec:ud}

In this section, we study the metric perturbations up to second order in the uniform density gauge, i.e.,
\begin{eqnarray}
  {\rm{d}} s^2 & = & a^2 \left( - (1 + 2 \phi) {\rm{d}} \eta^2 + 2 \partial_i b
  {\rm{d}} \eta {\rm{d}} x^i + \left( (1 - 2 \psi) \delta_{i   j} +
  \frac{1}{2} h_{i   j} \right) {\rm{d}} x^i {\rm{d}} x^j \right)\ . 
\end{eqnarray}
The energy momentum tensor up to first order is given by
\be
  T_{00} & = & a^2 (1 + 2 \phi) \rho^{(0)}\ , \\
  T_{0 i} & = & - a^2 (P^{(0)} + \rho^{(0)}) \upsilon_i^{(1)} - a^2 \rho^{(0)}
  \partial_i b\ , \\
  T_{i   j} & = & a^2 \delta_{i   j} ((1 - 2 \psi) P^{(0)} +
  P^{(1)})\ . 
\ee
The density perturbations $\rho^{(1)}$ are zero in the uniform density gauge. 

Similar to discussions in the previous section, the equations of motion of the first order scalar perturbations are given by 
\begin{eqnarray}
  3\mathcal{H}^2 \phi + 3\mathcal{H} \partial_0 \psi +\mathcal{H} \Delta b -
  \Delta \psi & = & 0\ , \\
  \psi - \phi - \(2\mathcal{H}+ \partial_0\) b & = & 0\ , \\
  \partial_0^2 \psi +\mathcal{H} \left( \partial_0 \phi + 3 \partial_0 \psi +
  \frac{1}{3} \Delta b \right) - \frac{1}{3} \Delta \psi & = & 0\ . 
\end{eqnarray}
In the momentum space, we rearrange the above three equations as follows
\begin{eqnarray}
  \phi_k + x \partial_x \psi_k - \frac{x}{3} (k   b_k) + \frac{x^2}{3}
  \psi_k & = & 0\ , \label{35}\\
  - \frac{1}{x^2} \partial_x \(x^2 (k   b_k)\) - \phi_k + \psi_k & = & 0\ ,
  \\
  - \phi_k + x \partial_x \phi_k + 2 x \partial_x \psi_k + x^2 \partial_x^2
  \psi_k & = & 0\ .  \label{37}
\end{eqnarray}
By solving Eqs.~(\ref{35})--(\ref{37}), we obtain the transfer functions of $\psi_k$, $\phi_k$, and $k
  b_k$ to be
\begin{eqnarray}
  T_{\phi} (x) & = & - \frac{\sqrt{3}}{2} x   \sin \left(
  \frac{x}{\sqrt{3}} \right), \\
  T_{\psi} (x) & = & - \frac{3}{2} \cos \left( \frac{x}{\sqrt{3}} \right) +
  \frac{3 \sqrt{3}}{x} \sin \left( \frac{x}{\sqrt{3}} \right), \\
  T_b (x) & = & \frac{3}{2} x \left( 1 - \frac{6}{x^2} \right) \cos \left(
  \frac{x}{\sqrt{3}} \right) - 3 \sqrt{3} \left( 1 - \frac{3}{x^2} \right)
  \sin \left( \frac{x}{\sqrt{3}} \right) . 
\end{eqnarray}
Here, we notice that the amplitudes of $T_{\phi}(x)$ and $T_{b}(x)$ increase as $x$ when $x\gg 1$. 
In fact, as will be shown in this work, they lead to the discrepancy between the Newtonian gauge and the uniform density gauge. 

Based on the second order Einstein field equations, we obtain the source term at the right hand side of the Eq.~(\ref{6}), namely,
\begin{eqnarray}
  \Lambda^{i   j}_{l   m} S_{i   j} & = & \Lambda_{l
    m}^{i   j} \Big( 2 \phi \partial_i \partial_j \phi - \psi
  \partial_i \partial_j \psi + \frac{1}{\mathcal{H}^2} \partial_0 \psi
  \partial_i \partial_j \partial_0 \psi + \partial_s b \partial^s \partial_i
  \partial_j b + \Delta b \partial_i \partial_j b + \frac{2}{\mathcal{H}} \partial_0 \psi \partial_i \partial_j \phi \nonumber\\
  &  & + 2
  \psi \partial_i \partial_j \phi + 4\mathcal{H} \psi \partial_i \partial_j b
  + 3 \partial_0 \psi \partial_i \partial_j b + 2 \psi \partial_i \partial_j
  \partial_0 b  + 4\mathcal{H} \phi \partial_i \partial_j b + \partial_0 \phi
  \partial_i \partial_j b + 2 \phi \partial_i \partial_j \partial_0 b \Big)\ .  \label{47}
\end{eqnarray}
Following Eqs.~(\ref{9}) and (\ref{47}), we obtain the transfer function to be 
\begin{eqnarray}
  f \left( \left| \textbf{k} - \textbf{p} \right|, p, \eta \right) & = & 2
  T_{\phi} \left( \left| \textbf{k} - \textbf{p} \right| \eta \right) T_{\phi}
  (p \eta) - T_{\psi} \left( \left| \textbf{k} - \textbf{p} \right| \eta
  \right) T_{\psi} (p \eta) + \frac{1}{\mathcal{H}^2} T'_{\psi} \left( \left|
  \textbf{k} - \textbf{p} \right| \eta \right) T'_{\psi} (p \eta) \nonumber\\
  &  & - \frac{\left( \textbf{k} - \textbf{p} \right) \cdot
  \textbf{k}}{\left| \textbf{k} - \textbf{p} \right|   k}   T_b
  \left( \left| \textbf{k} - \textbf{p} \right| \eta \right) T_b (p \eta) +
  \frac{1}{\mathcal{H}} \left( T'_{\psi} \left( \left| \textbf{k} - \textbf{p}
  \right| \eta \right) T_{\phi} (p \eta) + T_{\phi} \left( \left| \textbf{k}
  - \textbf{p} \right| \eta \right) T'_{\psi} (p \eta) \right)\nonumber\\
  &  & + 2\mathcal{H} \Big( \frac{1}{p} T_{\psi} \left( \left| \textbf{k} -
  \textbf{p} \right| \eta \right) T_b (p \eta) + \frac{1}{\left| \textbf{k} -
  \textbf{p} \right|} T_b \left( \left| \textbf{k} - \textbf{p} \right| \eta
  \right) T_{\psi} (p \eta) + \frac{1}{p} T_{\phi} \left( \left| \textbf{k} -
  \textbf{p} \right| \eta \right) T_b (p \eta) \nonumber\\
  &  &+ \frac{1}{\left| \textbf{k} -
  \textbf{p} \right|} T_b \left( \left| \textbf{k} - \textbf{p} \right| \eta
  \right) T_{\phi} (p \eta) \Big) + \frac{1}{\left| \textbf{k} - \textbf{p} \right|} T_b' \left( \left|
  \textbf{k} - \textbf{p} \right| \eta \right) T_{\psi} (p \eta) +
  \frac{1}{2 p} T_{\phi} \left( \left| \textbf{k} - \textbf{p} \right| \eta
  \right) T'_b (p \eta)\nonumber\\
  &  & + \frac{1}{2 \left| \textbf{k} - \textbf{p} \right|} T'_b \left( \left|
  \textbf{k} - \textbf{p} \right| \eta \right) T_{\phi} (p \eta) +
  \frac{1}{p} T_{\phi} \left( \left| \textbf{k} - \textbf{p} \right| \eta
  \right) T'_b (p \eta) + \frac{1}{\left| \textbf{k} - \textbf{p} \right|} T'_b
  \left( \left| \textbf{k} - \textbf{p} \right| \eta \right) T_{\phi} (p
  \eta)\nonumber\\
  &  & + T_{\psi} \left( \left| \textbf{k} - \textbf{p} \right| \eta \right)
  T_{\phi} (p \eta) + T_{\phi} \left( \left| \textbf{k} - \textbf{p} \right|
  \eta \right) T_{\psi} (p \eta) + \frac{1}{p} T_{\psi} \left( \left|
  \textbf{k} - \textbf{p} \right| \eta \right) T'_b (p \eta) \nonumber\\
  &  & + \frac{3}{2} \left( \frac{1}{p} T'_{\psi} \left( \left| \textbf{k} -
  \textbf{p} \right| \eta \right) T_b (p \eta) + \frac{1}{\left| \textbf{k} -
  	\textbf{p} \right|} T_b \left( \left| \textbf{k} - \textbf{p} \right| \eta
  \right) T'_{\psi} (p \eta) \right)\ . \label{48}
\end{eqnarray}
For the counter term $\Lambda^{l m}_{i   j} \mathcal{X}_{l m}$, we obtain its kernel function by making use of Eq.~(\ref{15}), namely, 
\begin{eqnarray}
  I_{\chi} \left( \left| \textbf{k} - \textbf{p} \right|, p, \eta \right) & =
  & \frac{2}{\left| \textbf{k} - \textbf{p} \right| p} T_b \left( \left|
  \textbf{k} - \textbf{p} \right| \eta \right) T_b (p \eta)\ . \label{49}
\end{eqnarray}
Therefore, the kernel function of the gauge invariant gravitational waves in Eq.~(\ref{20}) is obtained to be
\begin{equation}
  I_H \left( \left| \textbf{k} - \textbf{p} \right|, p, \eta \right) =
  \frac{4}{k^2} \int^{\eta}_0 {\rm{d}} (k \bar{\eta}) \left\{ \frac{a
  (\bar{\eta})}{a (\eta)} k   G_{\textbf{k}} (\eta, \bar{\eta}) f
  \left( \left| \textbf{k} - \textbf{p} \right|, p, \bar{\eta} \right)
  \right\} - \frac{2}{\left| \textbf{k} - \textbf{p} \right| p} T_b \left(
  \left| \textbf{k} - \textbf{p} \right| \eta \right) T_b (p \eta)\ , 
\end{equation}
where $f \left( \left| \textbf{k} - \textbf{p} \right|, p, \bar{\eta} \right)$ is given in Eq.~(\ref{48}), and $k   G_k (\eta, \bar{\eta}) = \sin (k (\eta - \bar{\eta}))$ in the radiation dominated epoch. 
The counter term in the uniform density gauge is non-trivial.

\section{Comparison among the kernel functions}\label{sec:kernel}

\begin{figure}[h]
\includegraphics[width=1\textwidth]{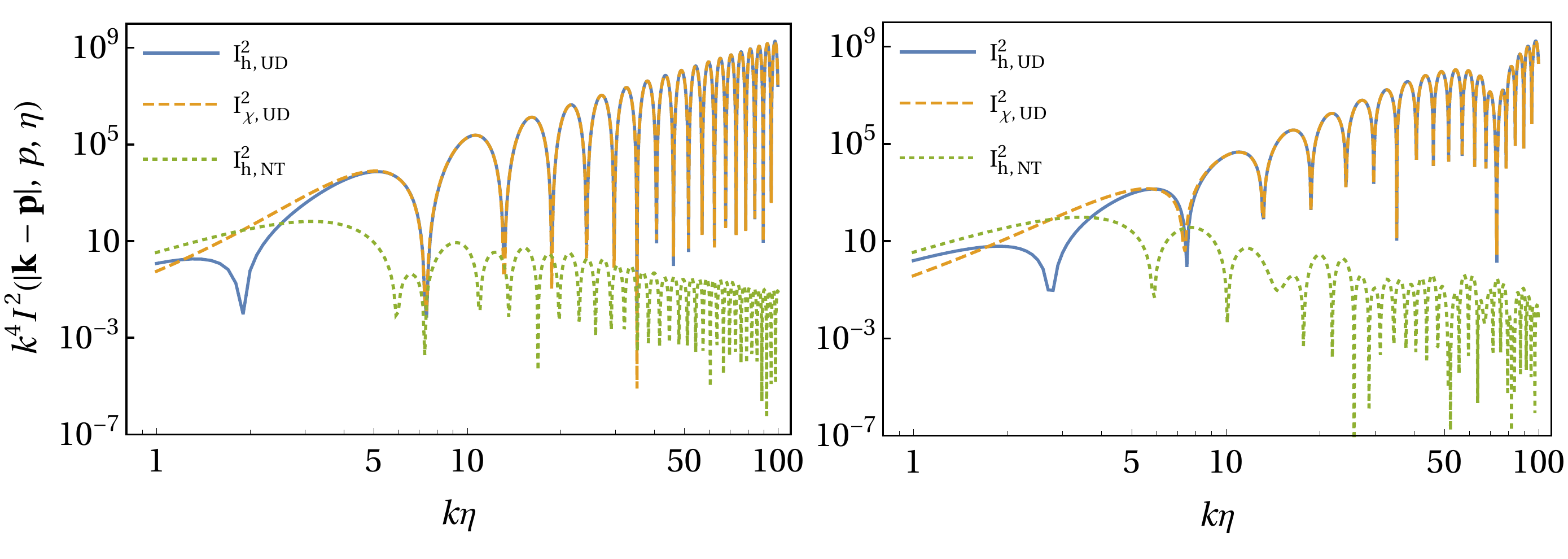}
\caption{The kernel functions $k^4 I^{2}_{h,\rm{NT}}(|\textbf{k}-\textbf{p}|,p, \eta)$ in the Newtonian gauge (dotted curve), $I^{2}_{h,\rm{UD}}(|\textbf{k}-\textbf{p}|,p,k \eta)$ in the uniform density gauge (solid curve), and $I^{2}_{\chi,\rm{UD}}(|\textbf{k}-\textbf{p}|,p,k \eta)$ for the counter term (dashed curve) in the uniform density gauge. \emph{Left panel}: we let $|\textbf{k}-\textbf{p}|=p=k$. \emph{Right panel}: we let $|\textbf{k}-\textbf{p}|=k$ and $p=0.1k$.}\label{F1}
\end{figure}

\begin{figure}[h]
	\includegraphics[width=1\textwidth]{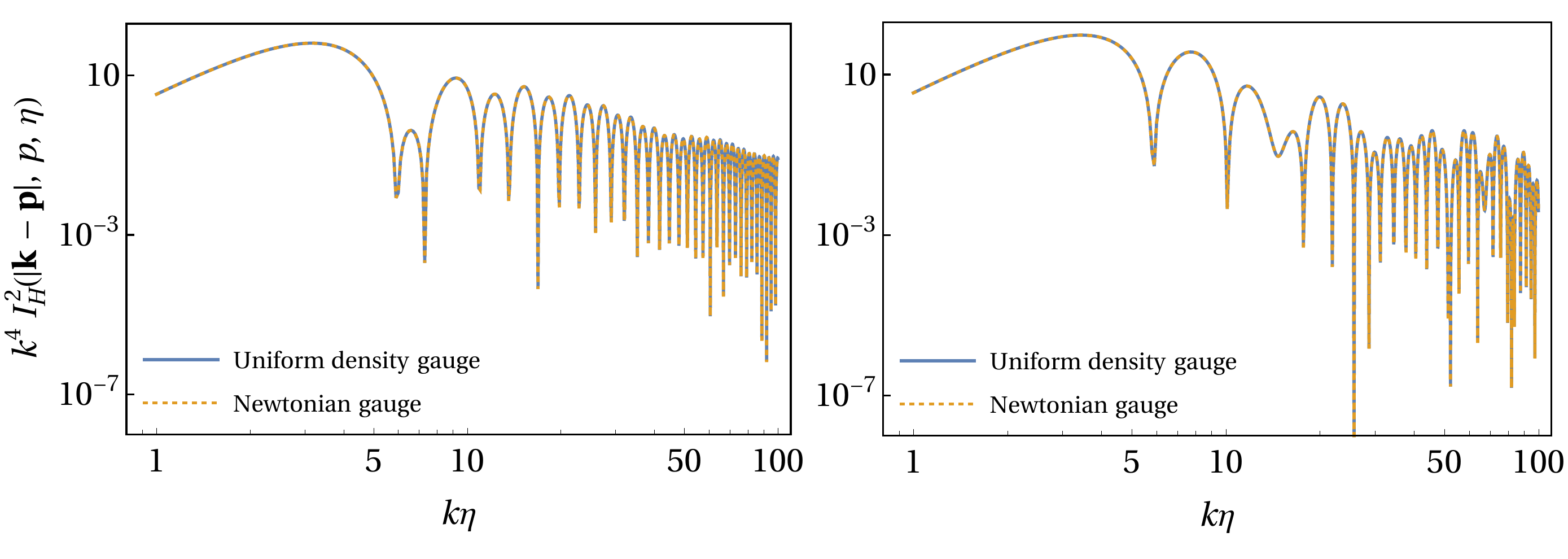}
	\caption{The gauge invariant kernel functions $k^4 I_H^{2}(|\textbf{k}-\textbf{p}|,p, \eta)$ in the Newtonian gauge (dotted curve) and the uniform density gauge (solid curve). \emph{Left panel}: we let $|\textbf{k}-\textbf{p}|=p=k$. \emph{Right panel}: we let $|\textbf{k}-\textbf{p}|=k$ and $p=0.1k$.}\label{F2}
\end{figure}

In this section, we compare the kernel functions in the Newtonian gauge and the uniform density gauge. 
We show that both of them lead to the same gauge invariant kernel function. 
Therefore, the energy density spectrum of the scalar induced gravitational waves should be the same in both of the two gauge fixings. 
To be specific, we show our results of the kernel functions in Fig.~\ref{F1} and Fig.~\ref{F2}. 

We compare the kernel functions in the Newtonian gauge and the uniform density gauge in Fig.~\ref{F1}. 
It shows that the kernel function of $h_{i j}$ in the uniform density gauge tends to be divergent as $\eta \rightarrow \infty$, while the one in the Newtonian gauge tends to converge. 
This result was also presented in the preview work \cite{Lu:2020diy}. 
It indicates that $h_{i j}$ is not gauge invariant, and the fictitious component of $h_{i j}$ should be subtracted. 
To subtract the fictitious perturbations, we introduce the counter term in Eq.~(\ref{3}). 
The counter term in the Newtonian gauge has been shown to be zero \cite{Chang:2020iji}. 
However, the counter term in the uniform density gauge is not trivial. 
In Fig.~\ref{F1}, we plot its kernel function, which is shown to be divergent as $\eta \rightarrow \infty$. 

In Fig.~\ref{F2}, we show the gauge invariant kernel functions $I_H^{2}(|\textbf{k}-\textbf{p}|,p, \eta)$ in both of the Newtonian gauge and the uniform density gauge. 
We find that they are the same for different gauge fixings. 
In this sense, they would lead to the same energy density spectrum of the scalar induced gravitational waves. 
Therefore, the gauge invariant induced gravitational waves is expected to be a well-defined observable.
In fact, we have shown this vital viewpoint in our previous works \cite{Chang:2020tji,Chang:2020iji}. 
As examples, we display the explicit derivations in two specific gauge fixings in the present work.

\section{Conclusion and discussion}\label{sec:conc}
In this work we have explicitly calculated the gauge invariant kernel functions, which uniquely determine the energy density spectrum of the scalar induced gravitational waves, in the Newtonian gauge and the uniform density gauge, respectively. 
Composed of the first order scalar perturbations, the counter term has been introduced in order to subtract the fictitious tensor perturbations. 
For the kernel functions, the discrepancy between the two gauge fixings has been shown to be explicitly eliminated in the gauge invariant framework. 
Therefore, we could obtain the same energy density spectrum for the scalar induced gravitational waves in both of the Newtonian and uniform density gauges. 
We have considered two typical gauge fixings as the specific examples of the gauge invariant framework. 
However, in fact, the gauge invariant method is also available to other gauge fixings considered in the previous works \cite{Inomata:2019yww,Yuan:2019fwv,DeLuca:2019ufz,Lu:2020diy,Tomikawa:2019tvi,Hwang:2017oxa,Giovannini:2020qta,Ali:2020sfw}. 
This conclusion is obvious, since in Eq.~(\ref{3}) we can fix an arbitrary gauge that does not change the gauge invariant kernel function quantitatively. 
Moreover, it is straightforward to generalize the gauge invariant method to study the induced gravitational waves in the matter dominated epoch.

\begin{acknowledgements}
We acknowledge Prof. Qing-Guo Huang, Mr. Zu-Cheng Chen, Mr. Chen Yuan, and Mr. Jing-Zhi Zhou for helpful discussions. 
This work is supported by the National Natural Science Foundation of China upon Grant No. 12075249, No. 11675182 and No. 11690022, and by a grant upon Grant No. Y954040101 from the Institute of High Energy Physics, Chinese Academy of Sciences.
We acknowledge the \texttt{xPand} package \cite{Pitrou:2013hga}. 
\end{acknowledgements}

\bibliography{igw-gauge-invariance}

\end{document}